\documentclass[%
onecollumn,
groupedaddress,notitlepage
 amsmath,amssymb,
 aps
]{revtex4-2}

\usepackage{amsmath}
\usepackage{graphicx}% Include figure files
\usepackage{dcolumn}% Align table columns on decimal point
\usepackage{bm}% bold math
\begin{document}

\title{Optical Extreme Learning
Machines with Atomic Vapors}% Force line breaks with \\

\author{Nuno A. Silva}
\email{nuno.a.silva@inesctec.pt}
\author{Vicente Rocha}%
\author{Tiago D. Ferreira}%
\affiliation{INESC TEC, Centre for Applied Photonics, Porto, Portugal}%

\date{\today}% It is always \today, today,

\begin{abstract}
Extreme learning machines explore nonlinear random projections to perform computing tasks on high-dimensional output spaces. Since training only occurs at the output layer, the approach has the potential to speed up the training process and the capacity to turn any physical system into a computing platform. Yet, requiring strong nonlinear dynamics, optical solutions operating at fast processing rates and low power can be hard to achieve with conventional nonlinear optical materials. In this context, this manuscript explores the possibility of using atomic gases in near-resonant conditions to implement an optical extreme learning machine leveraging their enhanced nonlinear optical properties. Our results suggest that these systems have the potential not only to work as an optical extreme learning machine but also to perform these computations at the few-photon level, paving opportunities for energy-efficient computing solutions.
\end{abstract}

\maketitle

%\tableofcontents

\section{Introduction}

In the last decades, the advent of artificial intelligence and neuromorphic architectures has completely reshaped the computing landscape. In particular, the logic procedures and arithmetic operations of the von Neumann paradigm may now be replaced by an alternative optimization-based approach. This establishes opportunities in two distinct directions. On one hand, on the side of algorithms, where rules and strategies are now autonomously inferred from data-driven processes\cite{NN_review}. On the other hand, on the side of the hardware, co-locating processing and memory operations pave for simpler hardware solutions capable of competing with electronic devices\cite{RC_review,RC_future,RC_polaritons}.

From the perspective of hardware, one of the most promising architectures is the reservoir computing framework. This design leverages the nonlinear dynamics of a physical system to simplify the transference of neuromorphic concepts to hardware implementations, allowing most physical systems to act as a computing platform. A particular case of a reservoir computer that neither uses temporal dependence nor system feedback are Extreme Learning Machines(ELM), an architecture closer to a common feed forward network and typically easier to implement\cite{ELM_huang2006, ELM_huang2012, ELM_Conti}. In short, an Extreme learning machine exploits an untrained nonlinear random transformation to project each element of an input space onto a high-dimensional output space. The learning process is then performed in this output space, i.e. at the output layer, which reduces the computational load and bypasses the need to fine-tune all the network weights\cite{ELM_huang2006}.

ELMs and reservoir computers have the potential to simplify the deployment of alternative physical systems as effective computing platforms, from mechanical\cite{RC_mechanical, RC_pendulum} and hydraulic\cite{RC_first, RC_fluids_vortex, RC_fluids_solitons,RC_waterwaves} to optical systems\cite{ELM_photonic, ELM_photonic_multiplexing, ELM_INESC,Neuromorphic_SiliconPhotonic}, being the latter the main focus of this work. Yet, deploying an effective physical ELM requires control of the relevant nonlinear dynamics\cite{ELM_Conti,ELM_nonlinear,RC_solitons}. From the optical computing perspective, this means that one needs optical media with a tunable nonlinear optical response. Nonlinear crystals or thermo-optical media may be explored for that purpose, but their relatively weak nonlinearity may limit their applications, requiring either large propagation distances that forbid miniaturization and increase losses, or higher laser power that can be prohibitive from the power efficiency perspective.

In this context, atomic gases in near-resonant conditions may constitute a valuable alternative for the construction of such devices. Indeed, provided with a suitable level structure, one can engineer the light-matter interaction conditions to achieve a tunable optical media with enhanced nonlinear susceptibilities at ultra-low intensities\cite{RO_lowpower,RO_review,RO_enhanced3level,RO_4level_liquid2}. Motivated by this background, this manuscript explores how such near-resonant media can be utilized for deploying effective optical extreme learning machines (OELM). For that, we will start by modeling the optical response of a typical $N$-type four-level atomic system to demonstrate how it can be utilized to obtain a highly nonlinear optical medium that is suitable to perform at low-intensity levels. Then, introducing the basic concepts of ELMs, we demonstrate how such configuration can be envisioned and used as a processing solution by using a phase encoding scheme and interferometric processes occurring during the nonlinear optical propagation of an optical beam inside a gas cell. By performing beam propagation numerical simulations under realistic physical conditions, we test the computing capabilities of these systems, demonstrating the importance of controlling the nonlinear properties of the system.

%%%%%%%%%%%%%%%%%%%%%%%%%%%%%%%%%%%%%%%%%%%%%%%%%
%%%%%%%%%%%%%%%%%%%%%%%%%%%%%%%%%%%%%%%%%%%%%%%%%

\section{Propagation of an optical beam in an atomic media under near-resonant conditions}

\begin{figure}
\centering
\includegraphics[width=\linewidth]{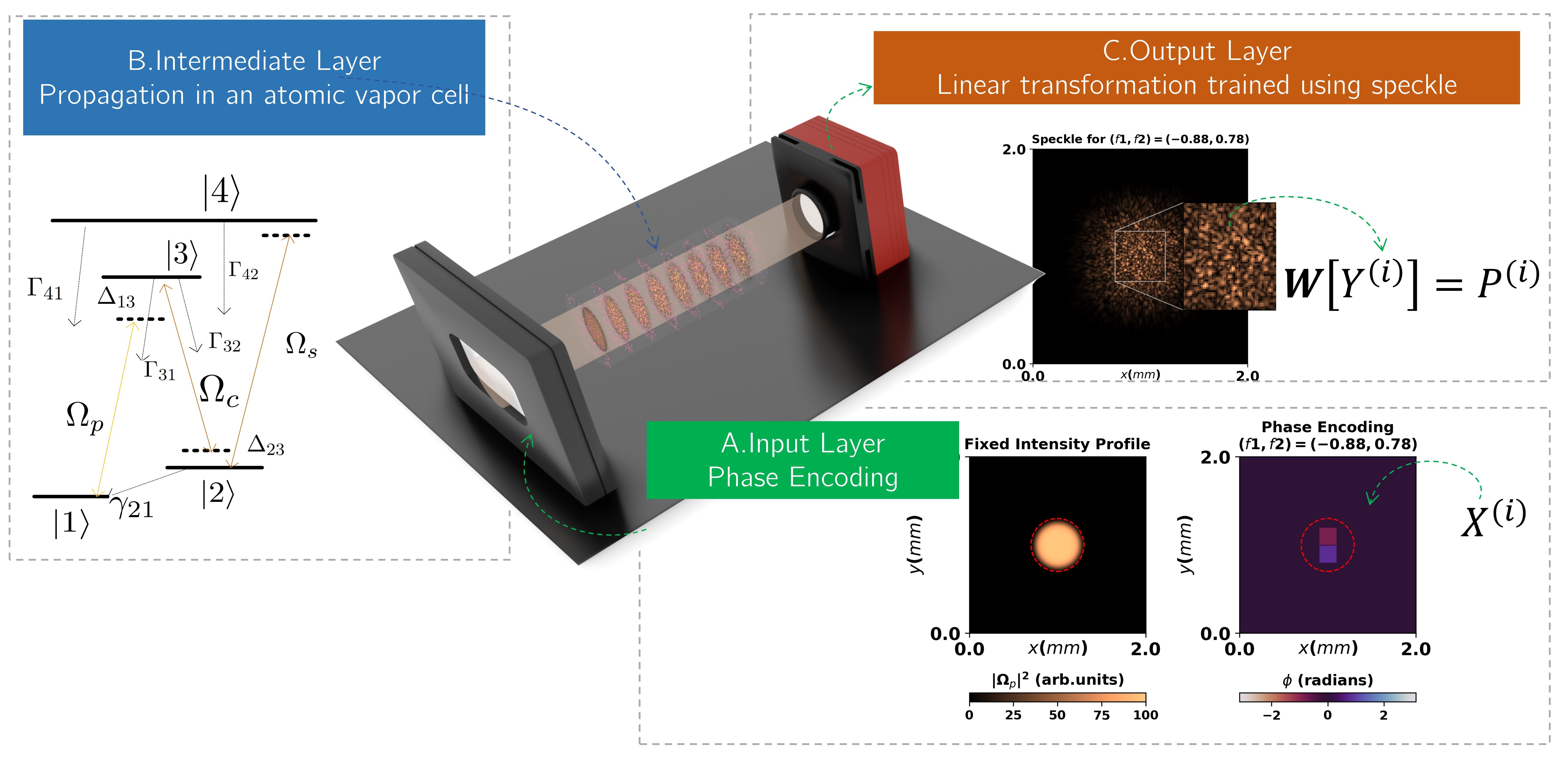}
\caption{Schematic of the purposed optical extreme learning configuration (exaggerated scales for illustration purposes). A. Each input of the dataset $\boldsymbol{X}^{(i)}$ is encoded into the phase profile of the incident flat top weak probe beam $\Omega_p$, which can be achieved with a spatial light modulator, for example. B. The probe beam propagates inside a cell filled with an atomic vapor with a $N$-type $4$-level atomic system, which acts as the intermediate layer of the OELM. C.After the propagation, the speckle pattern is recovered at the end, with the pixel intensity values being utilized as the output state $\boldsymbol{Y}^{(i)}$ for training the linear transformation $\boldsymbol{W}$ that completes the extreme learning machine architecture.}
\label{fig:N-conf}
\end{figure}

Regarding the goal of a strong nonlinear optical response, multiple atomic systems with a variety of level structures are known to support such phenomenology through effects related to quantum state coherence\cite{RO_review,RO_enhanced3level}. For this manuscript, we will focus on a typical 4-level atomic system\cite{RO_4level_liquid,RO_4level_liquid2,RO_N4level,QFL_persistent} interacting with three continuous-wave electromagnetic fields, which is widely known in the literature to support giant cross-Kerr nonlinearities even at ultra-low intensity levels\cite{RO_4level_liquid,RO_4level_liquid2}. In particular, we choose a typical $N$-type configuration(see Figure \ref{fig:N-conf}). First, a weak probe field $E_{p}=\frac{1}{2}\left[\mathcal{E}_{p}\left(r,z\right)e^{ik_{p}z-i\omega_{p}t}+\mbox{c.c.}\right]$, with envelope function $\mathcal{E}{}_{p}\left(r\right)$, center frequency
$\omega_{p}$ and wave vector $k_{p}$ couples
the levels $\left|1\right\rangle $ and  $\left|3\right\rangle $. The additional transitions are driven with stronger fields: a second ground state $\left|2\right\rangle $ is coupled to the excited state $\left|3\right\rangle $  via a control field $E_{c}=\frac{1}{2}\left[\mathcal{E}{}_{c}\left(r\right)e^{ik_{c}z-i\omega_{c}t}+\mbox{c.c.}\right]$, with envelope function $\mathcal{E}{}_{c}\left(r\right)$, center frequency
$\omega_{c}$ and wave vector $k_{c}$; and a switching field couples the second ground state $\left|2\right\rangle $ to a second excited state $\left|4\right\rangle $  via $E_{s}=\frac{1}{2}\left[\mathcal{E}{}_{s}\left(r\right)e^{ik_{s}z-i\omega_{s}t}+\mbox{c.c.}\right]$, with envelope function $\mathcal{E}{}_{s}\left(r\right)$, center frequency
$\omega_{s}$ and wave vector $k_{s}$.

Taking a semi-classical approach for the light-matter interaction\cite{NonlinearQuantumOptics, meystre2021quantum} and neglecting the effects of the weaker
probe beam on the dynamics of the control and switching fields, the propagation of the optical probe beam equation under the paraxial approximation is given by
\begin{equation}
ik_{p}\partial_{z}\mathcal{E}_{p}+\frac{1}{2}\nabla_{\perp}^{2}\mathcal{E}_{p}=\frac{1}{2\varepsilon_{0}c^{2}}\partial_{t}^{2}P_{p}.\label{eq:envelope-ev}
\end{equation} 
In this framework, the coherent light-matter interaction can be accounted for through the polarization $P_{p}$ density term that oscillates with $\omega_p$. For the current atomic medium it can be defined as $P_{p}=\eta\mu_{31}\rho_{31}e^{i\left(k_{p}z-\omega_{p}t\right)}+\mbox{c.c.}$, where $\mu_{ij}$ 
and $\rho_{ij}$ are the dipole moment
for the transition$\left|i\right\rangle \rightarrow\left|j\right\rangle $ and the population coherence terms of the density matrix operator $\rho$ respectively, and $\eta$ is the atomic density. 
To proceed with an analysis,
the dynamics of the atomic populations can be modeled by the master equation
\begin{equation}
\dot{\rho}=\frac{i}{\hbar}\left[\rho,\hat{H}\right]-\frac{\hat{\Gamma}\left(\rho\right)}{2},\label{eq:Master_equation}
\end{equation}
where $H$ is the system Hamiltonian given by
\begin{eqnarray}
\hat{H} & = & \sum_{i=1}^{4}\hbar\omega_{i}\left|i\right\rangle \left\langle i\right|-\hbar\left(\Omega_{p}e^{-i\omega_{p}t}\left|3\right\rangle \left\langle 1\right|+\right.\nonumber \\
 &  & \left.+\Omega_{c}e^{-i\omega_{c}t}\left|3\right\rangle \left\langle 2\right|+\Omega_{s}e^{-i\omega_{s}t}\left|4\right\rangle \left\langle 2\right|+\mbox{h.c.}\right),\label{eq:hamiltonian}
\end{eqnarray}
where the Lindblad
superoperator $\hat{\Gamma}\left(\rho\right)$ accounts for all the decoherence processes of the system, and the Rabi frequencies for the transitions 
are defined as $\Omega_{p,c,s}=\mu_{31,32,24}E_{p,c,s}/\hbar$. Using the definition of optical susceptibility 
$\chi_{p} = \frac{\eta\mu_{31}^{2}}{\varepsilon_{0}\hbar\Omega_{p}}\rho_{31}$, we can simplify the equation (\ref{eq:envelope-ev}) to
\begin{eqnarray}
i\frac{1}{k_{p}}\partial_{z}\Omega_{p}+\frac{1}{2k_{p}^{2}}\nabla_{\perp}^{2}\Omega_{p}+\chi_{p}\Omega_{p} & = & 0.\label{eq:Field probe equation}
\end{eqnarray}
Equations \ref{eq:Master_equation} and \ref{eq:Field probe equation} are then coupled through the susceptibility term, which may be obtained by solving the master equation. Assuming the steady-state solution, $\dot{\boldsymbol{\rho}}=0$ and by making use of the rotating-wave
approximation\cite{meystre2021quantum}, equation (\ref{eq:Master_equation})
can be expanded into the form
\begin{equation}
\dot{\boldsymbol{\rho}}=\left(M_{0}+M_{p}\left[\Omega _p\right]\right)\boldsymbol{\rho},\label{eq:density-ev}
\end{equation}
where $\boldsymbol{\rho}$ stands for a vectorized form of the 
$\rho_{ij}$ density matrix, and where $M_{0}$ and $M_{p}$
are matrices related with the $\Omega_{p}$-independent and dependent
parts of the master equation for $\boldsymbol{\rho}$, respectively. Recovering the weak probe beam assumption, i.e. $\left|\Omega_{p}\right|\ll\left|\Omega_{s}\right|,\left|\Omega_{c}\right|$, a perturbative approach to the weak probe beam gives $\rho_{ij}=\rho_{ij}^{(0)}+\rho_{ij}^{(1)}+\rho_{ij}^{(2)}+\rho_{ij}^{(3)}+\ldots$, obtained iteratively from
\begin{equation}
M_{0}\boldsymbol{\rho}^{(n)}=-M_{p}\boldsymbol{\rho}^{(n-1),}\label{eq:matrix-perturbation}
\end{equation}
starting from the ground state as the zero-th order solution (i.e.,
$\rho_{11}^{(0)}=1$, $\rho_{ij}^{(0)}=0$ if $i$ or $j\neq1$).

The results for this equation system are straightforward to obtain algebraically, but the general full expressions are typically too cumbersome\cite{QFL_persistent}. Yet, in the simplified limit of negligible dephasing processes between the two ground states, and assuming $\Gamma_{32}=\Gamma_{31}=\Gamma_{42}=\Gamma_{41}=\Gamma$, and the two-photon resonance condition $\Delta_{p}=\Delta_{c}=\Delta_{s}=\Delta$, it is possible to obtain that 
\begin{eqnarray}
\rho_{31}	&=&	\frac{2i\left|\Omega_{s}\right|^{2}}{\left(\Gamma-2i\Delta\right)\left(\left|\Omega_{c}\right|^{2}+\left|\Omega_{s}\right|^{2}\right)}\Omega_{p}-\frac{2i\left|\Omega_{s}\right|^{2}\left(\Gamma^{2}+4\Delta^{2}+16\left|\Omega_{s}\right|^{2}\right)\left|\Omega_{p}\right|^{2}}{\left(\Gamma+2i\Delta\right)\left(\Gamma-2i\Delta\right)^{2}\left(\left|\Omega_{c}\right|^{2}+\left|\Omega_{s}\right|^{2}\right)^2}\Omega_{p} \\
	&\approx&	-\frac{1}{2\Delta}\Omega_{p}+\frac{1}{4\Delta\left|\Omega_{0}\right|^{2}}\left|\Omega_{p}\right|^{2}\Omega_{p}
\end{eqnarray}
where the second approximation is valid for sufficiently large detunings, $\Delta\gg\Gamma,\Omega_{p,c,s}$ and equal amplitudes $|\Omega_c| = |\Omega_s| = |\Omega_0|$.
Finally, introducing the new variables $z'=k_{p}z$, $x'=k_{p}x$,
$y'=k_{p}y$, $r'=\sqrt{x'^{2}+y'^{2}}$, the transformation $\Omega_{p,s,c}^{'}=\Omega_{p,s,c}/\gamma$,
$\Delta_{1,2}^{'}=\Delta_{1,2}/\gamma$ and the coefficient $\kappa=\eta\mu_{31}^{2}/(\varepsilon_{0}\hbar\gamma)$
and dropping the primes, we obtain a dimensionless Nonlinear
Schr\"{o}dinger equation (NSE) to describe the evolution of the probe field as
\begin{equation}
i\partial_{z}\Omega_{p}+\frac{1}{2}\nabla_{\perp}^{2}\Omega_{p}+n\Omega_{p}-g\left|\Omega_{p}\right|^{2}\Omega_{p}=0,\label{eq:GNLSE}
\end{equation}
where the linear coefficient is given by
\begin{equation}
n=-\frac{1}{2\Delta},\label{eq:v_first}
\end{equation}
while the nonlinear term, associated with a self-Kerr effect, is
given by
\begin{equation}
g=-\frac{1}{4\Delta\left|\Omega_{0}\right|^{2}}.\label{eq:G_first}
\end{equation}

In the context of this work, the NSE model will be utilized to investigate numerically the propagation of a given envelope field $\Omega_p$ that contains the input information encoded in its wavefront. In the next section, we introduce how this propagation can be used to construct an optical computing system to process information by establishing a parallel to an extreme learning machine architecture. Furthermore, leveraging on the controllable parameters of the model, specifically the detuning $\Delta$, we investigate the impact of this choice in the overall performance of our OELM, showcasing the opportunities of using atomic systems for this specific purpose.

%%%%%%%%%%%%%%%%%%%%%%%%%%%%%%%%%%%%%%%%%%
%%%%%%%%%%%%%%%%%%%%%%%%%%%%%%%%%%%%%%%%%%

\section{Building an Optical Extreme Learning Machine}
To build an effective optical ELM we need first to understand the inner workings of this architecture. For this section, we consider the task of predicting a $\boldsymbol{P}^{\left(i\right)}$ belonging to a space $\mathbb{R}^{N_{target}}$ for a total of $N_D$ input states $\boldsymbol{X}^{\left(i\right)}$ belonging to the feature space $\mathbb{R}^{N_{input}}$ (with $N_{input}$ being the total number of features) and associated with ground-truth targets $\boldsymbol{T}^{\left(i\right)}$ that belong to a space $\mathbb{R}^{N_{target}}$.

The ELM architecture comprises three stages. First, each input state $\boldsymbol{X}^{\left(i\right)}$ 
is projected into an intermediate space as
\begin{equation}
\boldsymbol{Y}\left(\boldsymbol{X}^{\left(i\right)}\right)=\left[\begin{array}{c}
G\left(\boldsymbol{w}_{1}\boldsymbol{X},b_{1}\right)\\
\vdots\\
G\left(\boldsymbol{w}_{N_{c}}\boldsymbol{X},b_{N_{c}}\right)
\end{array}\right]
\label{eq:output}
\end{equation}
with $G$ being a nonlinear activation function, $N_{c}$ the number of output channels, and $\boldsymbol{w}_{i}$ and $b_{i}$ the internal weights and bias for each channel. The second step concerns the learning stage. It is performed on this high-dimensional intermediate space and assumes that we can train a linear transformation $\boldsymbol{W}$ from $\mathbb{R}^{N_{c}}$ to $\mathbb{R}^{N_{target}}$ and that matches the target values $\boldsymbol{T}$ by minimizing a given loss function. For example, it can be an output weight matrix $\boldsymbol{\Bar{W}}=\left[\boldsymbol{W}_{1}\cdots\boldsymbol{W}_{N_{c}}\right]^{T}$ that belongs to $\mathbb{R}^{N_{target}\times N_{c}}$ and that under the typical regularized ridge regression can be obtained from minimization of the loss function
\begin{equation}
    \boldsymbol{L}\left( \boldsymbol{\Bar{W}} \right) = \min_{\beta} \left[\left\Vert \sum_{i}\boldsymbol{\Bar{W}}\boldsymbol{Y}^{(i)}-\boldsymbol{T}^{(i)}\right\Vert ^{2}+\alpha \left\Vert\boldsymbol{\Bar{W}}\right\Vert^{2} \right],
\end{equation}
considering all the inputs $i$ and associated targets $\boldsymbol{T}^{(i)}$ of the training dataset. Note that while the Ridge model is typically suitable for regression tasks, classification is also possible for example by converting binary class targets to positive/negative values and keeping the sign of the prediction. Besides, under the theory of extreme learning machines\cite{ELM_huang2006,ELM_Conti,ELM_huang2012}, it is known that the universal approximation capabilities of this framework require two conditions under the activation function: i)an infinitely differentiable nonlinear activation function $G$ and ii)a random distribution of weights $\boldsymbol{w}_i$\cite{ELM_huang2006}.

Established the architecture, we can now discuss how to implement an OELM based on the propagation of an optical beam in our nonlinear optical media. First, we embed each input state into the phase of a given probe beam, namely
\begin{equation}
    \Omega_{p} (\boldsymbol{X}^{(i)},z=0) = \sum_j A_j(x,y) e^{i \phi(\boldsymbol{X}^{(i)}_j)}
\end{equation}
where $\phi(X_i)$ are encoding function and $A_j$ intensity distributions working as embedding states. Then, we let the optical beam propagate inside the media, before recovering it at the end $z=z_L$, utilizing an intensity sensor (e.g. a camera) for that purpose. Then, taking the pixels as the output channels we get 
\begin{equation}
\boldsymbol{Y}\left(\boldsymbol{X}^{\left(i\right)}\right)=\left[\begin{array}{c}
\left|\Omega_{p}\left(\boldsymbol{X}^{\left(i\right)},z=z_{L}\right)\right|_{1}^{2}\\
\ldots\\
\left|\Omega_{p}\left(\boldsymbol{X}^{\left(i\right)},z=z_{L}\right)\right|_{N_{c}}^{2}
\end{array}\right]
\end{equation}
where a subscript for the right-hand side $j=\left\{ 1,\ldots,N_{c}\right\}$ may for example refer to a pixel position $(x,y) = (mod(j/Nx),int(j/Nx))$ for a sensor of $N_x \times N_y$ pixels. Establishing a parallel with the definition in equation \ref{eq:output} it is straightforward to conclude that our nonlinear activation function is provided by a mixed combination of multiple effects, namely the interference of waves and generated patterns when interrogated with an intensity sensor, and the nonlinear evolution of the optical state inside the nonlinear media. While the first by itself does not warrant the necessary conditions for an ELM with universal approximation capabilities\cite{ELM_Conti,ELM_INESC}, it is known that the second may provide them if the strength of the nonlinearity is sufficiently high\cite{ELM_Conti,RC_solitons}.

%%%%%%%%%%%%%%%%%%%%%%%%%%%%%%%%%%%%%%%%%%
%%%%%%%%%%%%%%%%%%%%%%%%%%%%%%%%%%%%%%%%%%
%%%%%%%%%%%%%%%%%%%%%%%%%%%%%%%%%%%%%%%%%%

\section{Results}

In this section, we present the results of numerical simulations of equation \ref{eq:GNLSE} to explore the performance of an OELM for regression and classification tasks. Nevertheless, to increase the relevance of the work and maintain a close connection with a real-world implementation - supporting its future experimental implementation in cold or hot atomic gases - we have chosen as the $N$-type configuration the well-explored hyperfine structure
of the D line of $^{87}\mbox{Rb}$, more precisely the levels $6S_{1/2}(F=1)$, $6S_{1/2}(F=2)$, $6P_{1/2}(F=2)$
and $6P_{3/2}(F=1)$\cite{Rubidium_data}. Realistic physical parameters\cite{Rubidium_data} for this system are:
$\mu_{13}\simeq2.11\times10^{-29}\mbox{Cm}$,
$\mu_{23}\simeq1.26\times10^{-29}\mbox{Cm}$ and $\mu_{24}\simeq1.79\times10^{-30}\mbox{Cm}$ for dipole matrix elements;
$\Gamma=36\times10^{6}\mbox{s}^{-1}$ and $\gamma_{21}\simeq10^{-8}\gamma$ for decay rates.
Also, we consider the wavelengths of the optical fields to be $\lambda_{p}\approx\lambda_{c}=795\mbox{nm}$
and $\lambda_{s}=780\mbox{nm}$ and a fixed atomic concentration
$\eta=10^{18}\mbox{m}^{-3}$.
For the present work, we focus on the self-focusing Kerr regime, meaning that we will restrict our analysis to positive detunings. For the simulations presented below, we utilized $max(|\Omega_p|)=1.5$  in adimensional units (approx. $10mW/cm^2$), $|\Omega_0| = 4$. We further explore the range of $[1,200]$ for $g$ meaning that $\Delta$ would in practice vary within a range $[3.2\Gamma,646.7\Gamma]$. Encoding states are flattops with waist $w=400\mu m$ and propagate for a total distance of $z=1cm$ inside the optical media.

\subsection{Regression of nonlinear functions}

\begin{figure}
\centering
\includegraphics[width=\linewidth]{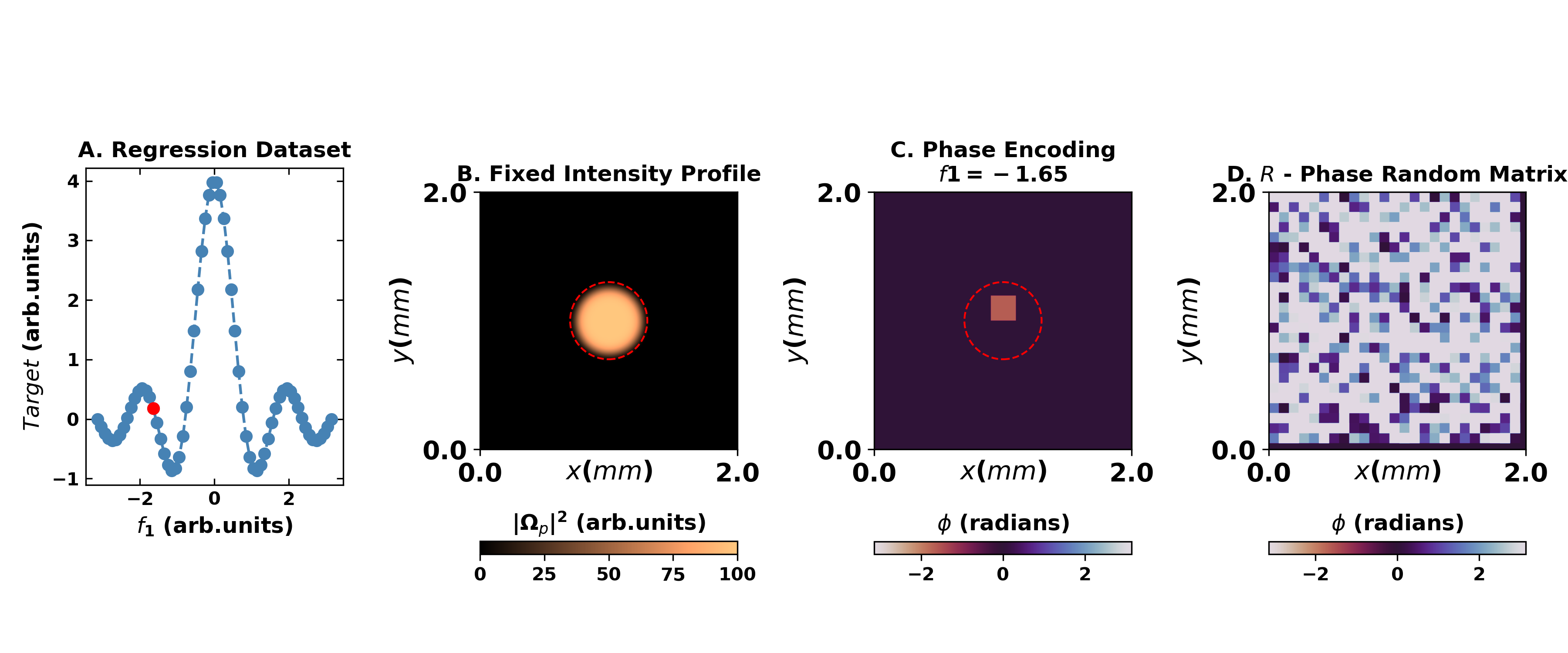}
\caption{Overview of the encoding for the regression task for dataset and targets presented in A. Panel B. presents the flattop intensity of the probe beam utilized, whereas C. presents one feature of the encoding scheme in the phase of the wavefront. D.A constant random matrix on the phase is also applied to the wavefront to warrant the speckle formation.}
\label{fig:regression_encoding}
\end{figure}

To understand the computing capabilities of our optical ELM we first focus on a typical regression task. For the purposes of this work, we have chosen to approximate the function $t(f_1)=sin(4f_1)/x$ with $f_1\in[-3,3]$ encoded in the input state as
\begin{equation}
    \Omega_{p} (\boldsymbol{X}^{(i)},z=0) = \left( A_0(x,y) + A_1(x,y) e^{i\phi(\boldsymbol{X}^{(i)}_1)} 
    \right)e^{iR(x,y)}
\end{equation}
with the encoding $\phi(\boldsymbol{X}^{(i)}_1) = 2\pi\frac{f_1-min(f_1)}{max(f_1)-min(f_1)} - \pi$ obtained using a min-max normalization of the features (see Figure \ref{fig:regression_encoding} for spatial distributions and encoding strategy). The additional fixed random phase distribution $R(x,y)$ warrants the generation of speckles in such a short distance and randomness of the projection onto the output space, and can be applied experimentally using a spatial light modulator or an optical diffuser, for example. For each state, we propagate it numerically, taking the intensity at $z=z_L$(i.e. imaging the intensity at the output plane) and recording a region of interest(ROI) of 60x60 pixels around the center $(x,y)=(0,0)$. We further downsampled the ROI by averaging regions of 2x2 pixels, before randomly choosing $N_c$ superpixels as the output state $\boldsymbol{Y}(\boldsymbol{X}^{(i)})$. Note that the effective dimensionality of the output space may be higher or lower than $N_c$ and it is associated with multiple factors such as the type of activation function and encoding strategy. Indeed, from a purely theoretical perspective, the only formal statement known is that if the matrix $\boldsymbol{H} = [\boldsymbol{Y}(\boldsymbol{X}^{(1)}), ..., \boldsymbol{Y}(\boldsymbol{X}^{(N_D)})]$ has $rank(H)=N_D$ then it can learn a dataset of $N_D$ elements with zero error, which would imply $N_D\leq N_c$\cite{ELM_huang2006}. Still, this statement does not impose major constraints on performance as one does not require $N_D\leq N_c$ for effective learning\cite{ELM_INESC,RC_solitons}. Indeed, a higher $N_c$ may be detrimental for performance as it may lead to overfitting issues, that must be dealt with using convenient regularization techniques.

\begin{figure}
\centering
\includegraphics[width=\linewidth]{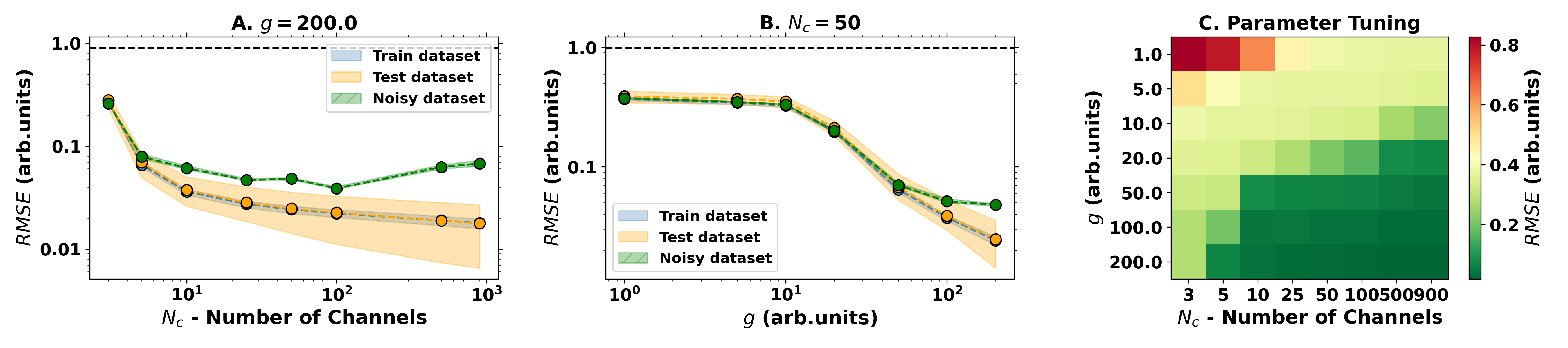}
\caption{Results for the regression task with the optical ELM measured with the root mean squared error (RMSE) metric. A. Change in performance with the increase of output channels $N_c$ with fixed $g=200$. B. Varying performance with the increase of the nonlinearity parameter $g$ for a fixed $N_c=50$. C.Performance on the test dataset with varying $g$ and $N_c$, showing a clear increasing performance tendency for stronger nonlinearities and larger output spaces.}
\label{fig:regression_results}
\end{figure}

To explore this framework we test the system with the regression of the function $t\left(f_1\right)$. The data set is comprised of 64 points equally distributed in the interval $\left[-3,3\right]$ with an $80-20\%$ train-test split where a standard Ridge Regression methodology from the \emph{sklearn} Python library is used to train the output layer\cite{sklearn}. Taking the root mean squared error (RMSE) as evaluation metric, the results obtained are presented in figure \ref{fig:regression_results}. As we can see, both the
the dimensionality of the output space $N_c$ and nonlinear strength $g$ are important for achieving a good performance in the regression task, achieving an error below 5\%, and meaning that both play a role as hyperparameters of the model. Yet, looking at subfigure \ref{fig:regression_results}C one can see that the nonlinear dynamics are more important for achieving
a good performance than simply increasing the number of output channels, meaning that the dimensionality is important but not sufficient. Indeed, this becomes evident when computing the error of the noisy dataset (same input data but with 5\% random noise added on top of it), for which one sees a decrease in the accuracy of the predictions with the increase of the $N_c$, which suggest an overfitting of the model with the increase of $N_c$.

\subsection{Classification of the Spiral Dataset}

\begin{figure}
\centering
\includegraphics[width=\linewidth]{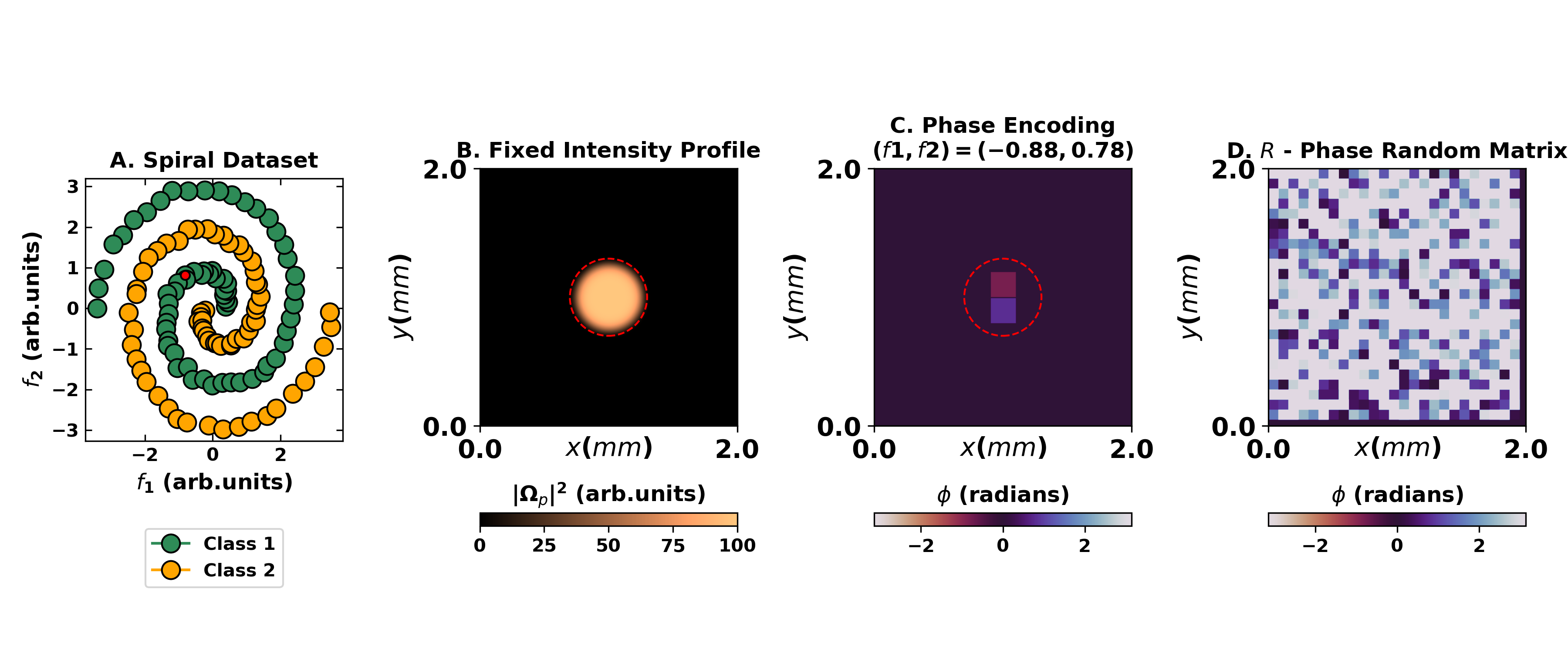}
\caption{Overview of the encoding for the classification task for dataset and targets presented in A. Panel B. presents the flattop intensity of the probe beam utilized, whereas C. presents a two-feature encoding scheme in the phase of the wavefront, for the point highlighted in red in panel A. D.A constant random matrix on the phase is also applied to the wavefront to warrant the speckle formation.}
\label{fig:spiral_encoding}
\end{figure}

For the second case study, we focus on the typical two-class spiral dataset to get a grasp of the classification capabilities of our proposed implementation as well as to get a clear picture of its generalization capabilities. Taking the dataset represented in Figure \ref{fig:spiral_encoding}, we encoded each pair of features $(f_1,f_2)$ into an input state
\begin{equation}
    \Omega_{p} (\boldsymbol{X}^{(i)},z=0) = \left( A_0(x,y) + A_1(x,y) e^{i\phi(\boldsymbol{X}^{(i)}_1)} + 
    A_2(x,y) e^{i\phi(\boldsymbol{X}^{(i)}_2)}
    \right)e^{iR(x,y)}
\end{equation}
with the encodings $\phi(\boldsymbol{X}^{(i)}_i) = 2\pi\frac{f_i-min(f_i)}{max(f_i)-min(f_i)}-\pi$ obtained using a min-max normalization of the features (see Figure \ref{fig:spiral_encoding} for spatial distributions). The associated vector on the output space for each input state was computed following the same strategy as utilized for the regression task.

\begin{figure}
\centering
\includegraphics[width=\linewidth]{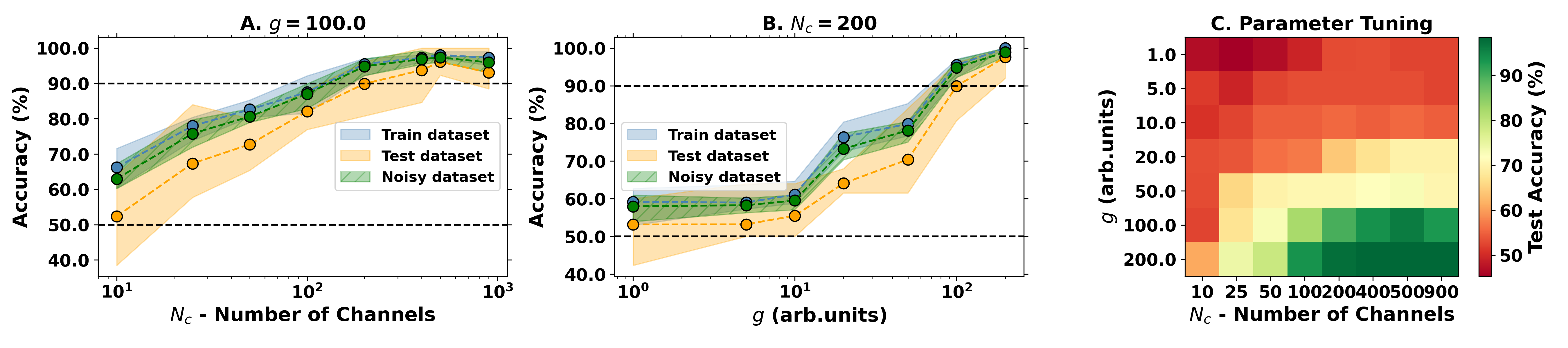}
\caption{Results for the classification task with the optical ELM. A. Change in performance with the increase of output channels $N_c$ with fixed $g=100$. B. Varying performance with the increase of the nonlinearity parameter $g$ for a fixed $N_c=50$. C. Performance on the test dataset with varying $g$ and $N_c$, showing a clear increasing performance tendency for stronger nonlinearities and larger output spaces.}
\label{fig:spiral_results}
\end{figure}

Utilizing the logistic regression as a classification model, we followed the same 80\%–20\% train-test subset division procedure, and varied the number of channels $N_c$ and the nonlinear strength $g$, obtaining the results depicted in Figure \ref{fig:spiral_results}. As in the regression task, the results again suggest that the output dimensionality and nonlinear strength of the physical system are both important parameters, but only the nonlinear strength warrants good accuracy. In ideal conditions, train and test accuracies above 90\% can be achieved, meaning that the model is not overfitting and validating our conceptual proposal as an effective OELM capable of performing nonlinear classification tasks.

\begin{figure}
\centering
\includegraphics[width=\linewidth]{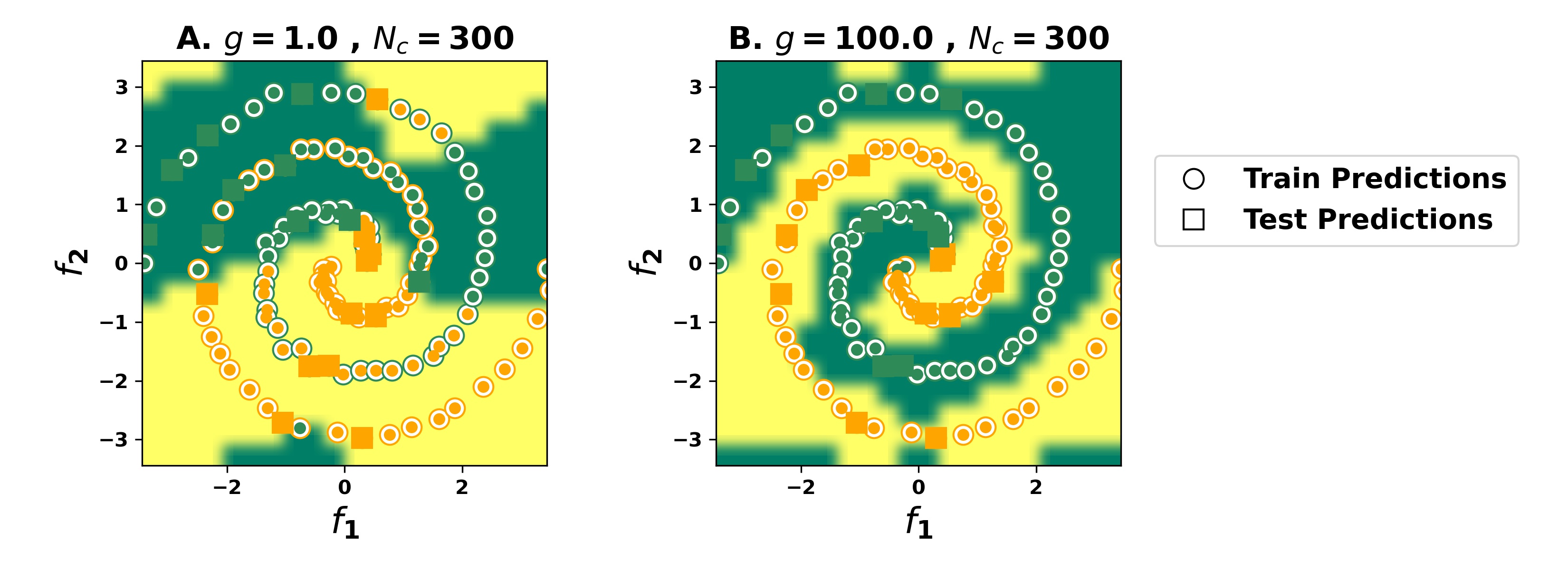}
\caption{Results for the two-spiral classification task with the optical ELM, regarding the generalization capabilities of the model for two distinct nonlinear parameters, A.$g=1.0$ and B.$g=100$.}
\label{fig:spiral_generalization}
\end{figure}

To finalize, Figure \ref{fig:spiral_results} also presents the accuracy for predictions on a noisy dataset, i.e. propagating states with 5\% random noise added on top of the initial state. As the performance does not vary significantly, it suggests that the classification model is robust and may generalize well for unseen data. Indeed, a better insight into model generalization capabilities may be achieved by performing numerical
simulations for each point covering a $20\times20$ rectangular grid with $f_1 \in \left[- \pi , \pi\right]$ and
$f_2 \in \left[- \pi , \pi\right]$, and predicting the associated label at each point using the previously trained model for $N_c=300$.
Figure \ref{fig:spiral_generalization} depicts the obtained results, showing that a correct spiral-like
separation of the data is possible for the sufficiently high nonlinear regime of the reservoir, demonstrating that generalization is possible but tightly connected with the strength of the nonlinearity.

%%%%%%%%%%%%%%%%%%%%%%%%%%%%%%%%%%%%%%%%%%

%%%%%%%%%%%%%%%%%%%%%%%%%%%%%%%%%%%%%%%%%%

\section{Discussion and concluding remarks}

The present manuscript reports a strategy to enable a physical implementation of an OELM using the nonlinear optical properties of atomic vapors in near-resonant conditions. The proposed implementation focuses on the use of an N-type configuration and the propagation of a weak probe beam assisted with strong coupling and switching fields. Following a perturbative approach, we derived an effective model for the propagation of a weak probe optical beam in the form of a nonlinear Schrödinger equation. Then, leveraging an encoding strategy based on the spatial modulation of the phase of the input probe beam, we established a connection between the physical system and the ELM architecture, demonstrating how one can benefit from the strong nonlinear optical properties of near-resonant optical media to enable an optical implementation of an ELM. Besides, by offering the possibility to control the nonlinearity strength with external parameters such as detuning or field intensity, the system presents an interesting playground to explore the crossover between linear and nonlinear response and to assess its impact on the performance of an optical ELM.  The numerical results presented demonstrate how combining a sufficiently large output dimensional space with strong nonlinear dynamics performs regression and classification of nonlinear problems. To approximate experimental conditions, realistic physical values together with synthetic noise are used to obtain the numerical results. Therefore, it is plausible to expect similar observations in experimental setups, which are soon to explored.

Finally, putting the results in perspective by comparing it with previous approaches in free space\cite{ELM_INESC,ELM_photonic}, the system presented here benefits from the fact that the nonlinearity does not reside solely in the measurement of the intensity of the field at the output plane (commonly done with an electronic element such as a camera) but also on the propagation itself, which can be controlled externally. These findings pave an important step for all-optical computing schemes and for establishing atomic vapors as possible building blocks of fast and robust neuromorphic all-optical computers.

\begin{acknowledgments}
This work is financed by National Funds through
the Portuguese funding agency, FCT – Fundação
para a Ciência e a Tecnologia, within the project
UIDB/50014/2020. T.D.F. is supported by FCT – Fundação
para a Ciência e a Tecnologia through Grant No.
SFRH/BD/145119/2019.
\end{acknowledgments}

\bibliography{bib}

\end{document}